\documentclass[openacc]{rstransa}

\usepackage{graphicx}
\usepackage{amssymb}
\usepackage{subfigure}
\usepackage{stackrel}
\usepackage{enumitem}
\usepackage{amsmath,leftidx}
\usepackage{amsfonts}
\usepackage{bm}
\usepackage{verbatim} 
\usepackage[normalem]{ulem}
\usepackage{soul}
\usepackage{mathtools}

\newcommand{\ket}[1]{\ensuremath{\left|{#1}\right\rangle}}
\newcommand{\bra}[1]{\ensuremath{\left\langle{#1}\right |}}

\newcommand{\beq}{\begin{equation}}
\newcommand{\eeq}{\end{equation}}
\newcommand{\bse}{\begin{subequations}}
\newcommand{\ese}{\end{subequations}}\newcommand{\bea}{\begin{eqnarray}}
\newcommand{\eea}{\end{eqnarray}}
\newcommand{\bit}{\begin{itemize}}
\newcommand{\eit}{\end{itemize}}
\newcommand{\bpmatrix}{\begin{pmatrix}}
\newcommand{\epmatrix}{\end{pmatrix}}

\newcommand{\be}{\begin{equation}}
\newcommand{\ee}{\end{equation}}
\newcommand{\ben}{\begin{eqnarray}}
\newcommand{\een}{\end{eqnarray}}

\begin{document}

\title{Indistinguishable entangled fermions: basics and future challenges}

\author{A. P. Majtey$^1$, A. Vald\'es-Hern\'andez$^2$, and E. Cuestas$^{1,3}$.}

\address{$^1$Instituto de F\'{i}sica Enrique Gaviola, CONICET and Universidad 
Nacional de C\'{o}rdoba,
Ciudad Universitaria, X5016LAE, C\'{o}rdoba, Argentina.\\
$^2$ Instituto de F\'{\i}sica, Universidad Nacional Aut\'{o}noma de M\'{e}xico, Apartado Postal 20-364, Ciudad de M\'{e}xico, Mexico.\\
$^3$ Quantum Systems Unit, OIST Graduate University, Onna, Okinawa 904-0495, Japan.
}

\subject{Quantum Physics}


\corres{Ana Majtey\\
\email{anamajtey@unc.edu.ar}}

\begin{abstract}
\small{The study of entanglement in systems composed of identical particles raises interesting challenges with far-reaching implications in both, our fundamental understanding of the physics of composite quantum systems, and our capability of exploiting quantum indistinguishability as a resource in quantum informa-tion theory. Impressive theoretical and experimental advances have been made in the last decades that bring us closer to a deeper comprehension and to a better control of entanglement. Yet, when it involves composites of indistinguishable quantum systems, the very meaning of entanglement, and hence its characterization, still finds controversy and lacks a widely accepted definition. 
The aim of the present paper is to introduce, within an accessible and self-contained exposition, the basic ideas behind one of the approaches advanced towards the construction of a coherent definition of entanglement in systems of indistinguishable particles, with focus on fermionic systems. 
We also inquire whether the corresponding tools developed for studying entanglement in identical-fermion systems can be exploited when analysing correlations in distinguishable-party systems, in which the complete information of the individual parts is not available. Further, we open the discussion on the broader problem of constructing a suitable framework that accommodates entanglement in presence of generalized statistics.
}
\end{abstract}

\maketitle
\section{Introduction}

While classical indistinguishability between identical particles is associated with the limited abilities of the experimenter to perfectly distinguish them, in the quantum realm particles of the same species are fundamentally indistinguishable \cite{Schrodinger1950}.
Though exchanging two such particles produces no change in the composite system's physical properties, the corresponding wave function may at most change sign (acquire a specific global phase) depending on the bosonic or fermionic nature of the individuals. In the former case the wave function remains the same, whence it is symmetric under the exchange of particles, whereas in the fermionic case it changes sign, so it is antisymmetric under the exchange \cite{sakurai_book}. Such properties of the bosonic and fermionic wave functions carries dramatic consequences on the statistical behaviour of the composites.

Another major difference between the classical and the quantum domains manifests itself in correlated composite systems. Classical systems have accessible states that are all mutually exclusive ---so no coherent superposition of distinct states is physically realizable---, whereas in quantum systems the physical possibilities are infinitely increased as the superposition principle holds. These features limit the kind of correlations that may arise in classical systems, in contrast to those that may be present in quantum composites \cite{Adesso_2016}. The firstly recognized non-classical correlation \cite{epr1935, Schrodinger_1935} was eventually known as \emph{entanglement} \cite{amico_2008,horodecki_2009,nielsen_chuang_book}. It is considered a paradigmatic quantum phenomenon that stresses the concept of locality \cite{
bell_1964, Wiseman2007}, and constitutes one of the fundamental resources for technologies based on quantum information \cite{Jozsa2003}.

The archetype of quantum information protocols involve individually addressable ---hence distinguishable--- parties.
The rapidly increasing development of quantum information science was therefore accompanied by an extensive study of the physical notion and mathematical characterization of entanglement in composites of distinguishable entities. 
However, by definition, the ideal of individually-addressable parties is intrinsically unattainable when dealing with \emph{indistinguishable}-particle systems 
\footnote{This is the case when dealing with identical particles in a physical situation which prevents us from distinguishing them. Of course when the parties, though of identical species, may be distinguished ---as for example when the supports of the single-particle wave functions are basically centered around sufficiently separated spatial regions---, the identical-particle system does not conform an indistinguishable-party system, individual access to the particles is feasible and quantum tasks can be performed as is usually done for distinguishable subsystems.}.
Moreover, the standard formal definition of entanglement (by standard we mean applicable to distinguishable subsystems), together with the symmetrization postulate, indicated in particular that systems of indistinguishable fermions could simply not exist in a disentangled state.
The question of how entanglement should be understood and characterized in situations in which the identicalness of the particles must be taken into consideration  became thus of particular importance,
and has led to a revision of the concept of entanglement, resulting in diverse ---sometimes discrepant--- definitions of it in systems composed of identical bosons or fermions \cite{Benatti_2020}. 
In addition to the fundamental interest in extending the notion of entanglement in identical-particle systems, the emergence of quantum technologies that rely on entanglement and involve systems composed of indistinguishable particles ---for instance Bose-Einstein condensates, quantum dots, ultracold gases--- indicates that a deeper understanding of their entanglement properties is on high demand \cite{Morris_2020,becher_2020}.
 
The aim of the present paper is to briefly introduce the non-specialized reader to one of the approaches that advance in the direction of constructing a coherent definition of entanglement in systems of identical particles, particularly fermions \cite{schielmann_2001,Eckert_2002,ghirardi_2004,plastino_2009_epl,tichy_2011_JPB,majtey_2016,bouvrie_2017_b}\footnote{The entanglement between identical bosons requires a different analysis that will not be considered here. We refer the interested reader to Refs. \cite{Dalton_Identical_Bosons_I,Dalton_Identical_Bosons_II,tichy_2011_JPB}.}.
Although there are several approaches to deal with the problem (see e.g. \cite{Benatti_2020}) here we focus on one of them; specifically,
we will adhere to the idea that   
a generic quantum state is non-entangled whenever a complete set of physical properties can be attributed to all individual subsystems \cite{Ghirardi_2002}, i.e., if a projective measurement on each subsystem exists that leads to a result that can be predicted with certainty. 
This \emph{physical} notion of "possessing a complete set of properties" is fully compatible with the formal \emph{mathematical} definition of non-entangled states when dealing with distinguishable parties, and prompts the definition of entanglement between identical fermions that will be the focus of our discussion. 
We further inquire about the possibility of applying the formalism, in particular the entanglement criteria, that ensues from this new definition of entanglement to analyze correlations in distinguishable-party systems in which the complete information of the subsystems is not available. 
Finally, we briefly comment on the general problem of shaping an appropriate conceptual and mathematical framework for describing entanglement in the more vast scenario that includes fractional statistics, as a generalization of the statistics that correspond either to identical-bosonic or fermionic particles.


\section{Description of a quantum state}

\subsection{The density operator}

When describing the state of a quantum system in terms of elements of an appropriate Hilbert space $\mathcal{H}$, we resort to the notion of the \emph{state vector} $\ket{\psi}\in\mathcal{H}$ as the mathematical entity that represents the specific state of the system, and bears complete information regarding its physical attributes\footnote{Unless otherwise stated, throughout the exposition we will consider only \emph{normalized} state vectors, such that $\langle\psi|\psi\rangle=1$.}. 
Formally, for all physical variables $\{A_i\}$ whose associated operators (observables) $\{\hat A_i\} $ form a complete set of commuting observables (CSCO) it holds that 
\beq
\label{eigenvalue}
\hat A_i \ket{\psi}=a_i\ket{\psi},
\eeq
with $\ket{\psi}$ a vector of the eigenbasis common to all $\{\hat A_i\}$. Physically, Eq.  
(\ref{eigenvalue}) means that in the state $\ket{\psi}$ the variable $A_i$ has a well-defined (non-fluctuating, dispersionless) value, equal to $a_i$.

Therefore, the complete determination of the CSCO allows to determine the specific physical properties of the state, which are encoded in the corresponding set of quantum numbers that completely characterize the vector $\ket{\psi}$
\footnote{Physical variables $\{C_k\}$ for which the eigenvalue equation $\hat C_k\ket{\psi}=c_k\ket{\psi}$ does not hold are necessarily dispersive in the state $\ket{\psi}$, and consequently are not well-defined in such quantum state.} \cite{QMCOhen}. 

The description of the system's state in terms of a state vector therefore implies \emph{certainty} about a set of some physical variables, as for example when atoms are prepared in an eigenstate $\ket{E_n}$ of the Hamiltonian $\hat H$. 
However, not all physical situations fit well into this kind of description, and in many (most) of the cases we only have incomplete, purely statistical information of the specific state of the system (hence of all its physical variables), as for example when atoms are in thermodynamic equilibrium at temperature $T$, so the probability of finding them in a state of energy $E_n$ is proportional to $e^{-E_n/k_BT}$, with $k_B$ being the Boltzmann constant \cite{QMCOhen}.
To surmount the limitations inherent to the state vector description, we resort to the concept of \emph{density operator} as a mean to extend the description of a quantum state.

Consider a physical situation in which the quantum system of interest can be found in the state $\ket{\psi_1}$ (so it possesses all the physical properties attributable to $\ket{\psi_1}$) with probability $p_1$, in the state $\ket{\psi_2}$ with probability $p_2$, and so on, with $0\leq p_n\leq 1$ and $\sum_n p_n=1$. 
The mean value of the physical variable $A$ in such state will thus be $\langle \hat A\rangle=\sum_n p_n \langle \hat A\rangle_n$, with $\langle \hat A\rangle_n\equiv\langle \psi_n|\hat A|\psi_n\rangle$ the mean value of $A$ in the $n$-th state, hence 
\ben\label{meanA}
\langle \hat A\rangle&=&\sum_n p_n \langle \psi_n|\hat A|\psi_n\rangle\nonumber\\
&=&\sum_n p_n \textrm{Tr}\,(\ket{\psi_n}\!\bra{\psi_n}\hat A)\nonumber\\
\langle \hat A\rangle&=&\textrm{Tr}\,(\hat \rho \hat A),
\een
with
\beq \label{defrho}
\hat \rho=\sum_n p_n \ket{\psi_n}\!\bra{\psi_n}.
\eeq
Equation (\ref{meanA}) indicates that it is precisely the operator $\hat \rho$, known as the \emph{density operator}, the one that encodes all the information regarding the state of the system, consistently with the statistical information at disposal. 

Unlike the description that relies on the state vector $\ket{\psi}$, the formalism offered by the density operator allows for the study of physical systems for which there is only probabilistic information, introduced via the statistical weights $p_n$ that conform the probability distribution $\{p_n\}$.
When $p_n=\delta_{nk}$ for some $k$, Eq. (\ref{defrho}) reduces to 
\beq\label{pure}
\hat \rho=\ket{\psi_k}\!\bra{\psi_k},
\eeq
implying the certainty that the system's physical properties are those encoded in the state vector $\ket{\psi_k}$. 
Therefore, the density operator formalism accommodates the particular situation described by a single state vector, in which case it is said that $\hat \rho$ is a \emph{pure state}. For probability distributions other than $\{p_n=\delta_{nk}\}$, $\hat \rho$ is a statistical mixture, called a \emph{mixed state}. 

The most general quantum state is thus represented by a density operator $\hat \rho$ that pertains to the convex set $\mathcal{D}(\mathcal{H})$ of operators acting on $\mathcal{H}$. Due to the normalization of the states $\ket{\psi_n}$, it follows from Eq. (\ref{defrho}) that $\textrm{Tr}\,\hat \rho=\sum_n p_n\langle\psi_n|\psi_n\rangle=\sum_n p_n=1$, therefore a legitimate $\hat \rho$ has trace equal to 1. Further, it is also a positive semi-definite operator, which means that $\langle\phi|\hat \rho|\phi\rangle\geq 0$ for all $\ket{\phi}\in\mathcal H$ (whence is an Hermitian operator). These properties of the density operator ultimately ensue from the fundamental requirement that  the probabilities of finding the system in any given state are non-negative real numbers that add up to one.

As explained above, the decomposition (\ref{defrho}) reflects the fact that the system of interest can be found in the state $\ket{\psi_n}$ with probability $p_n$. However, there are other, in fact infinitely many different statistical ensembles that lead to the same predictions regarding the physical properties of the system. 
That is, for a given $\hat \rho$, there are infinite different sets $\{\pi_j;|\phi_j\rangle\}$ of probabilities $\pi_j$ and pure states $|\phi_j\rangle$ in which $\hat \rho$ can be decomposed in the form of (\ref{defrho}). These sets are related to each other by unitary transformations, and constitute the \emph{pure-state decompositions} of $\hat \rho$ \cite{nielsen_chuang_book}. 
Though all these decompositions are equally valid and give rise to the same density operator, there is a particular one that will be of practical use below, namely the \emph{spectral decomposition}:
\beq\label{spectral}
\hat \rho=\sum_{i}\lambda_i\ket{u_i}\!\bra{u_i},
\eeq
where $\{\ket{u_i}\}$ and $\{\lambda_i\}$ stand, respectively, for the  (orthonormal) eigenvectors of 
$\hat \rho$ and the corresponding eigenvalues, satisfying $0\leq \lambda_i\leq 1$, and $\sum_i\lambda_i=1$. 

\subsection{State's information content}

It follows from the previous discussion that a pure state implies maximal information about the state of the system (a set of physical variables are completely well-determined), whereas in a mixed state some degree of uncertainty exists about all the physical variables of the system. 
A natural connection thus arises between the nature of the density operator (whether it is pure or mixed) and the information content of the state. In order to formalize such connection we focus on the \emph{purity} of the state $\hat \rho$, defined as $\textrm{Tr}\,\hat \rho^2$. 
The purity is invariant under unitary transformations, whence it is the same for all the (infinitely many) pure-state decompositions. In particular, we can resort to (\ref{spectral}) to compute
\beq \label{purity}
\textrm{Tr}\,\hat \rho^2=\sum_{i}\lambda_i^2\leq1.
\eeq
The inequality in (\ref{purity}) saturates ---hence $\textrm{Tr}\,\hat \rho^2$ is maximal and equal to $1$---, if and only if $\lambda_i=\delta_{ij}$ for some $j$, i.e., if and only if $\hat \rho$ corresponds to a pure state so there is maximal information of the system's properties. 
On the other extreme, the minimum value of $\textrm{Tr}\,\hat \rho^2$ corresponds to  $\lambda_i=\lambda=1/D$, where $D=\dim{\mathcal H}$, that is, when all the states $\ket{u_i}$ in the decomposition (\ref{spectral}) are equally probable, corresponding to the worst scenario in terms of the information regarding the specific state of the system. In the latter case $\hat \rho$ is said to be a \emph{maximally mixed} state. We can therefore employ the difference
\beq\label{SL}
S_L(\hat \rho)=1-\textrm{Tr}\,\hat \rho^2
\eeq
to quantify the degree of uncertainty associated to the state $\hat \rho$, via the degree of `mixedness' of the density operator. 
The quantity $S_L(\hat \rho)$ defined in Eq. (\ref{SL}) is the \emph{linear entropy} of the state $\hat \rho$. 

\section{Correlated composite systems}\label{correlated}

Assume that $\hat \rho$ represents the state of a system $S$, which is a composite of two subsystems $S_1$ and $S_2$, i.e $S=S_1 +S_2$.
The mean value of an observable $\hat A_1$ corresponding to a physical variable of $S_1$ is, following (\ref{meanA}),
\beq
\label{meanA1}
\langle\hat A_1\rangle=\textrm{Tr}\;[\hat \rho\, (\hat A_1\otimes\mathbb I_2)]=\textrm{Tr}\;(\hat \rho_1\hat A_1),
\eeq
with
\beq
\label{defrho1}
\hat \rho_1=\sum_{i} \leftidx{_2}\!\langle i\,|\hat \rho\,|i\rangle_2=\textrm{Tr}_2\,\hat \rho,
\eeq
where $\{\ket{i}_2\}$ is an arbitrary basis that expands the Hilbert space associated to $S_2$, and $\textrm{Tr}_2$ denotes the operation of partial trace over $S_2$. 
The operator $\hat \rho_1$ is a legitimate density operator for subsystem $S_1$, and completely determines the mean value of any of its observables, as follows from (\ref{meanA1}).
Accordingly, (\ref{defrho1}) is identified with \emph{the state of subsystem} $S_1$, the (reduced) density operator from which all the physical information of $S_1$ can be extracted. 

Now, if we can assign well-defined values to a set of variables of the composite $S_1+S_2$, meaning that $\hat \rho$ is a pure state ($\hat \rho=\ket{\psi}\!\bra{\psi}$), we may wonder about the physical variables that are perfectly defined on each of the subsystems. Inspection of the state of $S_1$ gives 
\ben\label{rho1}
\hat \rho_1 =\sum_i \prescript{}{2}\!\langle i|\psi\rangle\langle\psi|i\rangle_2=\sum_i|\tilde{\phi}_i\rangle\langle\tilde{\phi}_i|
= \sum_i\langle\tilde{\phi}_i|\tilde{\phi}_i\rangle\ket{\phi_i}\!\bra{\phi_i}, 
\een
where $|\tilde{\phi}_i\rangle\equiv \prescript{}{2}\!\langle i|\psi\rangle$ is a non-normalized vector in $\mathcal{H}_1$ (the Hilbert space associated with $S_1$), and $\ket{\phi_i}=|\tilde{\phi}_i\rangle/\sqrt{\langle\tilde{\phi}_i|\tilde{\phi}_i\rangle}$ the corresponding normalized state. Since $p_i\equiv\langle\tilde{\phi}_i|\tilde{\phi}_i\rangle$ satisfies $0\leq p_i\leq1$, and $\sum_ip_i=1$ (as a consequence of $\hat \rho_1$ having trace equal to one), Eq. (\ref{rho1}) writes as
\beq\label{rho1mix}
\hat \rho_1=\sum_ip_i\ket{\phi_i}\!\bra{\phi_i},
\eeq
so unless $p_i=\delta_{ij}$ for some $j$, no physical variable of $S_1$ can be predicted with certainty. An analogous reasoning leads to the same conclusion regarding the second subsystem. This means that the best knowledge of the whole (complete information of $S_1+S_2$ via the pure state $\ket{\psi}$) does not necessarily implies the best knowledge of the parts (incomplete information of $S_1$, encoded in the  statistical mixture $\hat \rho_1$). 

The loss of information when passing from the state of the global system to the (reduced) states of the subsystems reflects an underlying correlation between the parties.
Indeed, if (\ref{rho1mix}) was a pure state ($p_i=\delta_{ij}$), it would mean that tracing $\ket{\psi}\!\bra{\psi}$ over the degrees of freedom of $S_2$ leaves unaffected the available information of $S_1$, meaning that no information common to $S_1$ \emph{and} $S_2$ is contained in $\ket{\psi}$ (otherwise, when tracing over $S_2$ some information about $S_1$ would have been lost). 
Consequently, by having complete information about the whole system $S$ \emph{and} its parts $S_i$, we conclude that $S_1$ and $S_2$ are independent subsystems, and that the information content about each of them is encoded in a separate fashion in  $\ket{\psi}$. If, on the contrary, (\ref{rho1mix}) was not a pure state, then some global information ---common to both $S_1$ and $S_2$--- is contained in $\ket{\psi}$, the subsystems are not independent and correlations between them exist. 

The above discussion can be summarized stating that for a composite system $S=S_1+S_2$ in a pure state $\ket{\psi}$, the subsystems are correlated if and only if   
\beq\label{corr}
S_L(\hat \rho_i)=1-\textrm{Tr}\,\hat \rho_i^2>0,
\eeq
with $i=1,2$\footnote{For $S$ in a pure state $\ket{\psi}$, it holds that $S_L(\hat\rho_1)=S_L(\hat\rho_2)$, so the same amount of information is being lost when tracing $\ket{\psi}\!\bra{\psi}$ over either one of the subsystems. The identification is a mathematical consequence of the \emph{Schmidt decomposition}.}.
Equation (\ref{corr}) establishes a criterion to determine whether the parts, jointly found in a pure state, are correlated or not. 
\subsection{Exchange correlations}

All particles of the same species, like all electrons or all photons, possess the same intrinsic physical properties, and behave identically under the same physical circumstance\footnote{This is a very strong assumption that is justified \textit{a posteriori}, due to its success in the sense that it constitutes the cornerstone upon which physical theories, that allow us to understand and manipulate our natural environment, are constructed.}. 
Consequently, in an identical-particle system no observation can distinguish between states differing by the permutation of any pair of the identical constituents. 
At a quantum level, an additional fundamental principle is at work. It is supported by empirical evidence and asserts that  states $\ket{\psi}$ of composites of particles of the same species exhibit a particular type of symmetry under the exchange of any two particles \cite{Messiah-Greenberg,Ballentine, Girardeau-1965}: states describing systems of identical bosons are symmetric, whereas states that describe systems of  identical fermions are anti-symmetric. 
This \emph{symmetrization postulate} implies the Pauli exclusion principle \cite{pauli_1940, Marton2018}, and lies at the root of physical phenomena of utmost importance that ensue from quantum statistics \cite{yukalov_2011,Dalfovo_1999,greiner_2003}. In particular, the Pauli exclusion principle has been proven to be crucial for the stability of matter \cite{dyson_1967,lenard_1968}.

Consider, for example, a system composed of a pair of identical fermions, each of which has two degrees of freedom, a spatial and an internal one. Let $\{\ket{A},\ket{B}\}$ and $\{\ket{0},\ket{1}\}$, denote orthonormal eigenstates of the spatial and the internal degree of freedom, respectively. Each particle has thus four accessible states: $\ket{A,0},\ket{A,1},\ket{B,0}$, and $\ket{B,1}$. 
According to the symmetrization postulate, the situation in which one fermion is with certainty in the state $\ket{A,0}$ whereas the other one is in $\ket{B,1}$ is represented by the \emph{Slater determinant}\footnote{We resort to the simplified notation 
$\ket{\varphi}\ket{\phi}$ for the product $\ket{\varphi}\otimes \ket{\phi}$, where the first (second) vector refers always to one and the same particle.}
\begin{equation}\label{DslaterAB}
    \ket{\psi}^{\textrm{SD}}=\frac{1}{\sqrt{2}}(\ket{A,0}\ket{B,1}-\ket{B,1}\ket{A,0}) \,,
\end{equation}
and results from antisymmetrizing the product state
\beq\label{prodAB}
\ket{\psi}^{\textrm{prod}}=\ket{A,0}\ket{B,1},
\eeq
describing the same physical situation (one fermion in $\ket{A,0}$, and the other one in $\ket{B,1}$) but only accessible to a pair of \emph{distinguishable} fermions.
In $\ket{\psi}^{\textrm{prod}}$ it is clear that the particle occupying the state $\ket{A,0}$ is the first one; in contrast, when dealing with $\ket{\psi}^{\textrm{SD}}$ it is undetermined \emph{which} particle (the first or the second one) is found in such state. 
But this information is physically irrelevant when the parties are indistinguishable, so stating that one fermion is in $\ket{A}(\ket{B})$ and prepared in the internal state $\ket{0}(\ket{1})$ suffices to assign a complete set of properties to the individual parties when the composite is found in  $\ket{\psi}^{\textrm{SD}}$\cite{Ghirardi_2002}. Of course, such assignment is also possible when the system's state is $\ket{\psi}^{\textrm{prod}}$. 

The indistiguishability of the fermions in the state (\ref{DslaterAB}) is mathematically captured via the antisymmetrization operation 
\beq
\mathcal A:\ket{\psi}^{\textrm{prod}}\rightarrow \ket{\psi}^{\textrm{SD}},
\eeq
which leads to the non-factorizable structure of (\ref{DslaterAB}). 
Such structure prevents the (reduced) state of a single fermion from being a pure state. Indeed, direct calculation shows that the density operator of a single fermion ensuing from (\ref{DslaterAB}) is the mixed state
\beq \label{fDM}
\hat \rho^{\textrm{SD}}_\textrm{f}=\frac{1}{2}(\ket{A,0}\!\bra{A,0}+\ket{B,1}\!\bra{B,1}).
\eeq
It thus follows from the criterion (\ref{corr}) that  $\ket{\psi}^{\textrm{SD}}$ encodes correlations between the parties. These correlations are rooted at their indistinguishability, and correspond to the so-called \emph{exchange correlations}. 

Exchange correlations arise in pure states of $N$ identical fermions whenever the state can be identified (in some appropriate basis of the antisymmetric Hilbert space) with a single Slater determinant\footnote{Also called a pure state of Slater rank 1.}. 
 The corresponding reduced density matrices of subsystems made up of $M$ fermions, $\hat \rho^{\textrm{SD}}_M$ (with $1\leq M\leq N-1$), have a purity equal to \cite{majtey_2016}
 \beq \label{bound_rho_SD}
 \textrm{Tr}\,\hat \rho^{\textrm{SD}}_M=\binom{N}{M}^{-1}<1,
 \eeq
 indicating that correlations are present among all bipartitions of the system. Moreover, since the global state is a Slater determinant, such correlations are compatible with the possibility of assigning a complete set of properties to the individual fermions 
 (as occurs in the distinguishable-particle counterpart, involving product states).  

  
\subsection{Entanglement in identical-fermion systems}

\subsubsection{Defining fermionic entanglemnent}

Resorting to the previously considered pair of identical fermions, let us focus on the state
\begin{equation}\label{Fentangled}
    \ket{\psi}^{\textrm{non-SD}}=\frac{1}{2}(\ket{A,0}\ket{B,1}-\ket{B,1}\ket{A,0}+
    \ket{A,1}\ket{B,0}-\ket{B,0}\ket{A,1}
    ) \,,
\end{equation}
which is a balanced coherent superposition of two Slater determinants, namely $\frac{1}{\sqrt{2}}(\ket{A,0}\ket{B,1}-\ket{B,1}\ket{A,0})$ and $\frac{1}{\sqrt{2}}(\ket{A,1}\ket{B,0}-\ket{B,0}\ket{A,1})$. The single-fermion density operator corresponding to the global state (\ref{Fentangled}) reads 
\beq \label{fDMentangled}
\hat \rho^{\textrm{non-SD}}_\textrm{f}=\frac{1}{4}(\ket{A,0}\!\bra{A,0}+\ket{B,1}\!\bra{B,1}+\ket{A,1}\!\bra{A,1}+\ket{B,0}\!\bra{B,0}).
\eeq
This is a mixed state, evincing correlations among the fermions. However, direct calculation shows that the purity of $\hat \rho^{\textrm{non-SD}}_\textrm{f}$ is lower than the purity of $\hat \rho^{\textrm{SD}}_\textrm{f}$, or equivalently, has a greater amount of mixedness. In line with the discussion below Eq. (\ref{rho1mix}), this implies that the loss of information when tracing over one fermion is greater when the composite is found in the state $\ket{\psi}^{\textrm{non-SD}}$ than when it is in the state $\ket{\psi}^{\textrm{SD}}$. This feature reveals the presence of correlations \emph{beyond} exchange correlations, that is, of correlations that add to those ensuing solely from the antisymmetrization procedure. 

Inspection of (\ref{Fentangled}) shows that the emergence of correlations beyond those attributable to the particle-exchange is accompanied by the impossibility of assigning well-defined properties to any of the parties, in sharp contrasts with the case in which the global state is a single Slater determinant. 
Indeed, the structure of (\ref{Fentangled}) is incompatible with the identification of one particle being in any of the states $\ket{A,0},\ket{A,1},\ket{B,0},\ket{B,1}$. This lack of determinacy about the properties of the subsystems is also present in the distinguishable-particle counterpart of the state vector (\ref{Fentangled}), namely the non-product state 
\begin{equation}\label{entangled}
    \ket{\psi}^{\textrm{non-prod}}=\frac{1}{\sqrt{2}}(\ket{A,0}\ket{B,1}+
    \ket{A,1}\ket{B,0}
    ) \,,
\end{equation}
which can be obtained from (\ref{Fentangled}) by ``freezing'' the spatial degrees of freedom ---so that the location identifies the particles---, thus transforming the pair of indistinguishable fermions into a pair of distinguishable parties \cite{Eckert_2002,bouvrie_2017_b}. 
In the opposite way, the state (\ref{Fentangled}) can be obtained from (\ref{entangled}) by applying the antisymmetrization operation, 
\beq
\mathcal A:\ket{\psi}^{\textrm{non-prod}}\rightarrow \ket{\psi}^{\textrm{non-SD}}.
\eeq

The above observations  naturally introduce the notion of \emph{fermionic entanglement} in pure states of indistinguishable fermions according to the following principle: pure states incompatible with the existence of a set of well-defined properties of all the individual subsystems will be called entangled states. In the indistinguishable-fermion system, such entanglement captures the correlations between the parties \emph{on top} of the exchange correlations. 
On the other hand, states for which a set of well-defined properties of the particles exists will be regarded as non-entangled, and all the correlations present are due solely to the antisymmetry of the state. 
From a mathematical perspective, this notion of entanglement is encoded in states that cannot be expanded (regardless of the basis) as a single Slater determinant \cite{Eckert_2002,Ghirardi_2002,pauskaukas_2001} ---as e.g. the state (\ref{Fentangled})---, whereas non-entangled states admit a decomposition in terms of a single Slater determinant, as in (\ref{DslaterAB}). 

Notice that if the physical conditions are such that the identical fermions can be distinguished ---as for example by freezing some degree of freedom that allows to address the particles individually--- the entangled states are those that cannot be expanded (in any basis) as a product state (as e.g. the state (\ref{entangled})), while non-entangled ones admit a product decomposition, as in (\ref{prodAB}). 
Therefore, when the fermions become individually addressable the notion of fermionic entanglement reduces to the standard notion of entanglement (between distinguishable parties), mathematically rooted at the factorizability of the state vector. However, within the present approach, entanglement and non-factorizability cease to be equivalent concepts. The former is defined via the \emph{physical} impossibility of attributing well-defined variables to the parties, 
whereas the latter identifies the \emph{mathematical} impossibility of factoring the state vector. 

It should be stressed that the aforementioned concept of entanglement between indistinguishable fermions is by no means the only way to address the problem of defining entanglement between  identical particles. 
In fact, a widely used approach to characterize entanglement in indistinguishable-party systems focuses on the entanglement between  \emph{orthogonal modes}, i.e., between \emph{distinguishable} occupation states. 
This approach relies on the operationally motivated concept of entanglement as a correlation involving subsystems that can be measured (hence addressed) individually, and consequently requires them to be distinguishable \cite{Benatti_2020,Morris_2020}. 
The so-called \emph{mode entanglement} approach thus emphasizes the usefulness of entanglement in applications, and formally resembles the standard  entanglement definition when the global state is expressed in the Fock representation. Since from this perspective the physical particles are put aside in favour of the occupancies of different modes, their  indistinguishability is eluded  in the discussion. 
 

\subsubsection{Quantifying fermionic entanglement}
The previous analysis can be extended to composites of $N$ identical fermions. In particular, the presence of entanglement in the global state $\ket{\psi}$ can be certified according to the following criterion \cite{plastino_2009,majtey_2016}, based on the purity of the density operator $\hat \rho_M$ of a subsystem containing $M$ of the $N$ fermions (so $1 \leq M\leq N-1$):   
\begin{subequations}
\label{entcritBouvrie}
\begin{eqnarray}
&\textrm{Tr}\rho_M^2=\binom{N}{M}^{-1} \quad \Leftrightarrow\quad \ket{\psi}\text{ is non-entangled,}\\ &\frac{1}{d_M}\leq \textrm{Tr}\rho_M^2<\binom{N}{M}^{-1} \quad \Leftrightarrow\quad \ket{\psi}\text{ is entangled,} 
\end{eqnarray}
\end{subequations}
where $d_M=\binom{d}{\textrm{min}\{M,N-M\}}$, and $d$ stands for the dimension of the single-fermion Hilbert space. 
This criterion states that only non-entangled states correspond to reduced density matrices with maximal purity (consistent with the symmetry imposed on the global state), so a higher degree of mixedness evinces the presence of entanglement. Entanglement criteria based on the degree of mixedness of the reduced density matrices allows for the establishment of entanglement measures such as the  \emph{concurrence} for pure states of distinguishable q-dits \cite{Wootters_1997,Rungta_2001}, or the \emph{fermionic concurrence} for the identical-fermion system \cite{majtey_2016}. 
This latter measure rests on (\ref{entcritBouvrie}), holds for pure states of $N$ $d$-dimensional fermions,  vanishes whenever the state reduces to a single Slater determinant, and generalizes (for global pure states) the measure introduced in \cite{Eckert_2002}, originally restricted to the case $N=2$, $d=4$. 

Entropic-based entanglement measures for indistinguishable fermions rely in general on entanglement criteria that extend (usually in a straightforward mathematical way) those applicable to systems of distinguishable parties, to make them useful for detecting quantum correlations beyond exchange correlations \cite{schielmann_2001,pauskaukas_2001,manzano_2010,tichy_2011_JPB,Zander_2012,majtey_2016}.
Some of these measures reduce to the corresponding amount of entanglement for distinguishable parties whenever the particles becomes individually addressable \cite{bouvrie_2017_b} (see also \cite{Benatti_2020} for a critical discusion on this matter).

Thought we have focused only in global pure states, the entanglement measures in the identical-fermion system can be extended to mixed states via the pure-state decomposition of the density operator, or by means of the convex roof extension, as is done in the distinguishable-particle case \cite{Uhlmann2010,tichy_2011_JPB}.

\section{Discussion and outlook} 

A question that naturally arises from the various notions of entanglement in systems of identical particles is whether we can employ the formalism, tools or concepts developed for analysing entanglement in quantum systems of identical constituents, to deal with systems of distinguishable parties, but which we cannot access individually due to some lack of information of the subsystems (attributable to the experimental setup or to the particular description of the system).
To exemplify this we focus on  two  fermions of different species but equal mass $m$, in a  one-dimensional harmonic trap characterized by a frequency $\omega$. The fermions interact via a zero-range or delta potential, so  the Hamiltonian of the system reads
\begin{align}
\label{eq_H_full}
{ \hat H} \! = \! -\frac{\hbar^2}{2m}\!\left(\frac{\partial^2}{\partial x_1^2}+\frac{\partial^2}{\partial x_2^2}\right) \!+ 
\frac{m \omega^2}{2}\!\left(x_1^2 + x_2^2 \right)\!+ g \delta(x_1-x_2),
\end{align}

\noindent where $x_1$ and $x_2$ are the particles' positions and $g$ is the strength of the interaction.  
Relying on a very precise control of interactions and confinement, the authors of Ref. \cite{zurn_2012} showed that the fundamental state
of this system, in the limit of strong  repulsive interaction, coincides with that of a system of two non-interacting identical fermions, therefore experimentally confirming that \emph{fermionization} ---at first thought for bosonic particles  \cite{girardeau_1960}---, arises also for distinguishable particles. 
The term fermionization is due to Girardeau in 1960, and originally referred to a regime in which a bosonic system exhibits the same momentum distribution and density profile as a system of non-interacting fermions \cite{girardeau_1960}. 
The concept was later extended to distinguishable bosons and fermions, and was experimentally proven for fermions in 2012 \cite{girardeau_2010, zurn_2012, rontani_2012}. 
The observation that in the limit of infinite interaction the fundamental state of the pair of interacting distinguishable particles resembles a state of two identical non-interacting fermions can be understood as an effect of the hard-core interactions that induce an effective Pauli exclusion principle, even in the bosonic or distinguishable case. Therefore fermionization can be thought as a map between strongly interacting bosons or distinguishable particles and a gas of non-interacting fermions. In other words, in the strong repulsive regime the effect of the interaction term in the Hamiltonian (\ref{eq_H_full}) amounts to impose a constraint in the symmetric wave function: it must vanish whenever two particles are at the same position, thus resembling the Pauli exclusion principle for identical fermions.

The Hamiltonian (\ref{eq_H_full}) admits eigenfunctions that depend on the relative distance between the particles, $|u_1 - u_2|$ with $u_{1(2)} = \sqrt{m \omega / \hbar} \,x_{1(2)}$.  
Such eigenfunctions are symmetric in the variables $u_1$ and $u_2$ and antisymmetric in the variables $u_<$ and $u_>$, where $u_<$ ($u_>$) represents the spatial coordinate of the particle located to the left (right), or equivalently  $u_>=\textrm{max}\{u_1,u_2\}$ and $u_<=\textrm{min}\{u_1,u_2\}$. 
Since we are dealing with two distinguishable fermions, the symmetry of the eigenfunctions of $\hat H$ is due to the symmetry of the potential rather than to an \textit{ad hoc} imposition related to the nature of the particles. 
However, in Ref. \cite{cuestas_2020_ferm} it is argued that the solutions in terms of $u_<$ and $u_>$ account for a lack of information about the location of the individual particles: one of them is on the left and the other one on the right, but we cannot distinguish which one (1 or 2) lies at each side of the boundary $x_1=x_2$.    
In this case it seems reasonable to resort to the formalism employed when dealing with systems of indistinguishable fermions to extract information about the correlations between the particles with variables $u_{<,>}$.

In Ref. \cite{cuestas_2020_ferm} it was shown that the normalized eigenfunction for the ground state in the limit of infinite repulsive interaction can be written as
\begin{equation}
\label{phigsx1x2}
\psi_{\text{gs}}(u_1,u_2)=\frac{1}{\sqrt 2}\Big\vert\phi_0(u_1)\phi_1(u_2)-\phi_1(u_1)\phi_0(u_2)\Big\vert \,,
\end{equation}
%
where $\phi_{0(1)}(u)$ is the ground (first excited) wave function of the one-dimensional harmonic oscillator, $\phi_{0(1)}(u)=\mathcal{N}_{0(1)}e^{-u^2/2}H_{0(1)}(u)$,  with 
$\mathcal{N}_{0(1)}$ a normalization constant and  $H_{0(1)}$ the zeroth (first) Hermite polynomials. In terms of the variables $u_<, u_>$, Eq. (\ref{phigsx1x2}) expresses as a Slater determinant 
%
\begin{equation}
\label{phigsxmenorxmayor}
\psi_{\text{gs}}(u_<,u_>)=\frac{1}{\sqrt 2}\Big[\phi_0(u_<)\phi_1(u_>)-\phi_1(u_<)\phi_0(u_>)\Big].
\end{equation}
%

We can now resort to the state (\ref{phigsx1x2}) and to the criterion (\ref{corr}) to quantify the amount of correlations between the particles 1 and 2, obtaining  $S_L(\hat\rho_1)=1-\textrm{Tr}\hat\rho_1^2\sim0.36$ (here $\hat\rho_1$ stands for the reduced density matrix of particle 1).  
Analogously, from (\ref{phigsxmenorxmayor}) we compute the correlations between the (unidentified) particles located at $u_<, u_>$, and find $S_L(\hat \rho_<)=1-\textrm{Tr}\hat\rho_<^2 = 1/2$ ($\hat \rho_<$ being the reduced state of the left particle). 
In the first case, the correlation quantified by $S_L(\hat\rho_1)$ reflects entanglement between the (identifiable) particles 1 and 2, whereas in the second case the linear entropy $S_L(\hat\rho_<)$ accounts for the entanglement between the (unidentifiable) right and left particles. 
The results suggest that entanglement criteria developed for indistinguishable-fermion states can be of use here, in the sense that a vanishing fermionic entanglement for the state $\psi_{\text{gs}}(u_<,u_>)$ could act as a fermionization witness in laboratory. 
Indeed, the differentiation between the entanglement that arises from interactions and the entanglement rooted at the symmetry of the wave-function of two and three fermionic atoms, was recently addressed experimentally in Ref. \cite{becher_2020}.

Fermionization is a clear example that in certain physical circumstances  (which involve a strict one-dimensional motion and sufficiently strong interactions), bosons do not behave as bosons, but rather exhibit a fermionic behaviour. 
This evinces that when dealing with indistinguishable particles the binary outcome: Fermi-Dirac or Bose-Einstein statistics, may break down and generalized statistics emerge, enriching the physical possibilities of composites of identical quantum parties.
The incompleteness of the binary statistics scheme is even more evident if we take into account that when the interactions are continually changed from strongly attractive to strongly repulsive in an unpolarized Fermi gas (a system with the same amount of fermions of two distinguishable species) the system can be described as an ideal gas obeying \emph{generalized exclusion statistics} \cite{haldane_1991,wu_1994,guan_2007,shu_2010}, which allows for an intermediate behavior between the bosonic and the fermionic one, and even for more \textit{exclusive} fermions, relying on the base of \textit{any}onic quasiparticles. 
 
To the best of our knowledge the definition, characterization and quantification of entanglement in anyonic systems remains far from being understood. Most of the works made in the context of topological anyons focus on lattice systems and resort to the \emph{topological entanglement} by tracing over lattice sites or a space region\cite{levin_2006,kitaev_2006,bonderson_2017}, which recalls the notion of mode (not particle) entanglement. Other works within a many-body context address the entanglement in anyonic systems adhering to its standard definition (involving distinguishable parties), and do not discuss the inclusion (or exclusion) of the exchange correlations \cite{santachiara_2007}.
Another approach \cite{mani_2020} applies an information theoretic proposal  \cite{lofranco_2016,compagno_2018} to quantify the entanglement of two indistinguishable anyons in one dimension. 
Such proposal has the advantage of working without a fictitious labelling of the indistinguishable particles, hence is free of  an \textit{ad hoc} symmetrization procedure. 
What is symmetrized instead is the inner product, in such a way that it allows for an interpolation between the bosonic and the fermionic case, i.e. it naturally opens the possibility of considering more general statistics (yet, the proposal gives a mixture of mode and particle entanglement \cite{lourenco_2019}).
Even though all these works rely on different formalisms, the results consistently show that the entropy used as a measure of entanglement has a non-trivial dependence on the statistical parameter, the one that allows to continually transit from the bosonic to the fermionic statistics (see for instance Refs. \cite{santachiara_2007,mani_2020}). However, the nature of the connection between particle entanglement and the statistical parameter remains unclear. 

Fermionization ---and more generally the physics of anyonic particles--- does not only constitute a paradigmatic situation in which we need to extend and re-assess the existence of only two mutually exclusive statistics, but it also raises new questions that bring out the need for constructing a consistent broad and well-established theory for addressing entanglement in systems of indistinguishable particles. This is not a minor task, firstly because a unified language common to the many-body physics and the quantum information community is missing when attempting to discuss the many facets of quantum correlations, and secondly due to the lack of a consensus of what is actually meant by entanglement in identical-particle systems. 
Indeed, by focusing on different aspects of entanglement different notions of it emerge, which rest on its non-local properties, its operational meaning, its role as a quantum resource, or its relation to the (im)possibility of ascribing well-defined properties to the individual, entangled subsystems (which is the approach discussed here).
The differentiation between exchange correlations and legitimate entanglement depends on the adherence to one or another approach, and there is no universal or absolute definition of entanglement when the correlated parts become indistinguishable (yet of course most authors should agree in recognizing the presence of measurable correlations, irrespective of their specific identification). 
The formulation of widely (not cased-based) suitable definitions, and the consequent criteria of entanglement that apply consistently to composites of quantum particles irrespective of their anyonic, fermionic or bosonic character, stands as a major challenge in the search for a deeper comprehension of the quantumness of multiparticle systems. 

\vskip6pt

\enlargethispage{20pt}

\aucontribute{All authors equally performed the research, discussed the results and contributed in writing the paper. All authors have read and approved the final manuscript.}

\competing{The authors declare that they have no competing interests.}


\funding{A.P.M. and E.C. acknowledge funding from grant PICT 2017-2583 from ANPCyT (Argentina) and the Argentinian agencies SeCyT-UNC and CONICET for partial financial support. A.V.H. acknowledges financial support from DGAPA, UNAM through projects PAPIIT IN113720 and IN112723.}



\bibliographystyle{ieeetr}
\bibliography{bib_entanglement_IP}{}

\begin{thebibliography}{10}

\bibitem{Schrodinger1950}
E.~Schr\"{o}dinger, ``What is an elementary particle?,'' {\em Annual Report of
  the Board of Regents of The Smithsonian Institution}, pp.~183--196, 1950.

\bibitem{sakurai_book}
J.~J. Sakurai, {\em {Modern quantum mechanics; rev. ed.}}
\newblock Reading, MA: Addison-Wesley, 1994.

\bibitem{Adesso_2016}
G.~Adesso, T.~R. Bromley, and M.~Cianciaruso, ``Measures and applications of
  quantum correlations,'' {\em Journal of Physics A: Mathematical and
  Theoretical}, vol.~49, p.~473001, nov 2016.

\bibitem{epr1935}
A.~Einstein, B.~Podolsky, and N.~Rosen, ``Can quantum-mechanical description of
  physical reality be considered complete?,'' {\em Phys. Rev.}, vol.~47,
  pp.~777--780, May 1935.

\bibitem{Schrodinger_1935}
E.~{Schr{\"o}dinger}, ``{Die gegenw{\"a}rtige Situation in der
  Quantenmechanik},'' {\em Naturwissenschaften}, vol.~23, pp.~807--812, Nov.
  1935.

\bibitem{amico_2008}
L.~Amico, R.~Fazio, A.~Osterloh, and V.~Vedral, ``Entanglement in many-body
  systems,'' {\em Rev. Mod. Phys.}, vol.~80, pp.~517--576, May 2008.

\bibitem{horodecki_2009}
R.~Horodecki, P.~Horodecki, M.~Horodecki, and K.~Horodecki, ``Quantum
  entanglement,'' {\em Rev. Mod. Phys.}, vol.~81, pp.~865--942, Jun 2009.

\bibitem{nielsen_chuang_book}
M.~A. Nielsen and I.~L. Chuang, {\em Quantum Computation and Quantum
  Information}.
\newblock Cambridge University Press, 2000.

\bibitem{bell_1964}
J.~S. Bell, ``On the einstein podolsky rosen paradox,'' {\em Physics Physique
  Fizika}, vol.~1, pp.~195--200, Nov 1964.

\bibitem{Wiseman2007}
H.~M. Wiseman, S.~J. Jones, and A.~C. Doherty, ``Steering, entanglement,
  nonlocality, and the einstein-podolsky-rosen paradox,'' {\em Phys. Rev.
  Lett.}, vol.~98, p.~140402, Apr 2007.

\bibitem{Jozsa2003}
R.~{Jozsa} and N.~{Linden}, ``{On the role of entanglement in
  quantum-computational speed-up},'' {\em Proceedings of the Royal Society of
  London Series A}, vol.~459, pp.~2011--2032, Aug. 2003.

\bibitem{Benatti_2020}
F.~Benatti, R.~Floreanini, F.~Franchini, and U.~Marzolino, ``Entanglement in
  indistinguishable particle systems,'' {\em Physics Reports}, vol.~878,
  pp.~1--27, 2020.
\newblock Entanglement in indistinguishable particle systems.

\bibitem{Morris_2020}
B.~Morris, B.~Yadin, M.~Fadel, T.~Zibold, P.~Treutlein, and G.~Adesso,
  ``Entanglement between identical particles is a useful and consistent
  resource,'' {\em Phys. Rev. X}, vol.~10, p.~041012, Oct 2020.

\bibitem{becher_2020}
J.~H. Becher, E.~Sindici, R.~Klemt, S.~Jochim, A.~J. Daley, and P.~M. Preiss,
  ``Measurement of identical particle entanglement and the influence of
  antisymmetrization,'' {\em Phys. Rev. Lett.}, vol.~125, p.~180402, Oct 2020.

\bibitem{schielmann_2001}
J.~Schliemann, J.~I. Cirac, M.~Ku\ifmmode~\acute{s}\else \'{s}\fi{},
  M.~Lewenstein, and D.~Loss, ``Quantum correlations in two-fermion systems,''
  {\em Phys. Rev. A}, vol.~64, p.~022303, Jul 2001.

\bibitem{Eckert_2002}
K.~Eckert, J.~Schliemann, D.~Bruß, and M.~Lewenstein, ``Quantum correlations
  in systems of indistinguishable particles,'' {\em Annals of Physics},
  vol.~299, no.~1, pp.~88--127, 2002.

\bibitem{ghirardi_2004}
G.~Ghirardi and L.~Marinatto, ``General criterion for the entanglement of two
  indistinguishable particles,'' {\em Phys. Rev. A}, vol.~70, p.~012109, Jul
  2004.

\bibitem{plastino_2009_epl}
A.~R. Plastino, D.~Manzano, and J.~S. Dehesa, ``Separability criteria and
  entanglement measures for pure states of n identical fermions,'' {\em {EPL}
  (Europhysics Letters)}, vol.~86, p.~20005, apr 2009.

\bibitem{tichy_2011_JPB}
M.~C. {Tichy}, F.~{Mintert}, and A.~{Buchleitner}, ``{Essential entanglement
  for atomic and molecular physics},'' {\em Journal of Physics B Atomic
  Molecular Physics}, vol.~44, p.~192001, Oct. 2011.

\bibitem{majtey_2016}
A.~P. Majtey, P.~A. Bouvrie, A.~Vald\'es-Hern\'andez, and A.~R. Plastino,
  ``Multipartite concurrence for identical-fermion systems,'' {\em Phys. Rev.
  A}, vol.~93, p.~032335, 2016.

\bibitem{bouvrie_2017_b}
P.~A. Bouvrie, A.~Vald\'es-Hern\'andez, A.~P. Majtey, C.~Zander, and A.~R.
  Plastino, ``Entanglement generation through particle detection in systems of
  identical fermions,'' {\em Ann. Phys.}, vol.~383, p.~401, 2017.

\bibitem{Dalton_Identical_Bosons_I}
B.~M.~G. B.~J.~Dalton, J.~Goold and M.~D. Reid, ``Quantum entanglement for
  systems of identical bosons: I. general features,'' {\em Physica Scripta},
  vol.~92, no.~2, p.~023004, 2017.

\bibitem{Dalton_Identical_Bosons_II}
B.~M.~G. B.~J.~Dalton, J.~Goold and M.~D. Reid, ``Quantum entanglement for
  systems of identical bosons: Ii. spin squeezing and other entanglement
  tests,'' {\em Physica Scripta}, vol.~92, no.~2, p.~023005, 2017.

\bibitem{Ghirardi_2002}
G.~Ghirardi, L.~Marinatto, and T.~Weber, ``Entanglement and properties of
  composite quantum systems: A conceptual and mathematical analysis,'' {\em
  Journal of Statistical Physics}, vol.~108, pp.~49--122, 2002.

\bibitem{QMCOhen}
F.~L. Claude Cohen-Tannoudji, Bernard~Diu, {\em {Quantum Mechanics}}.
\newblock NYA: Wiley, 1977.

\bibitem{Messiah-Greenberg}
A.~M.~L. Messiah and O.~W. Greenberg, ``Symmetrization postulate and its
  experimental foundation,'' {\em Phys. Rev.}, vol.~136, pp.~B248--B267, Oct
  1964.

\bibitem{Ballentine}
L.~E. Ballentine, {\em Quantum Mechanics. A Modern Development}.
\newblock World Scientific, 1998.

\bibitem{Girardeau-1965}
M.~D. Girardeau, ``Permutation symmetry of many-particle wave functions,'' {\em
  Phys. Rev.}, vol.~139, pp.~B500--B508, Jul 1965.

\bibitem{pauli_1940}
W.~Pauli, ``The connection between spin and statistics,'' {\em Phys. Rev.},
  vol.~58, pp.~716--722, Oct 1940.

\bibitem{Marton2018}
J.~Marton, S.~Bartalucci, A.~Bassi, M.~Bazzi, S.~Bertolucci, C.~Berucci,
  M.~Bragadireanu, M.~Cargnelli, A.~Clozza, C.~Curceanu, L.~De~Paolis,
  S.~Di~Matteo, S.~Donadi, J.-P. Egger, C.~Guaraldo, M.~Iliescu,
  M.~Laubenstein, E.~Milotti, A.~Pichler, D.~Pietreanu, K.~Piscicchia,
  A.~Scordo, H.~Shi, D.~Sirghi, F.~Sirghi, L.~Sperandio, O.~Vazquez-Doce,
  E.~Widmann, and J.~Zmeskal, {\em Underground Test of Quantum Mechanics: The
  VIP2 Experiment}, pp.~155--168.
\newblock Cham: Springer International Publishing, 2018.

\bibitem{yukalov_2011}
V.~I. {Yukalov}, ``{Basics of Bose-Einstein condensation},'' {\em Physics of
  Particles and Nuclei}, vol.~42, pp.~460--513, May 2011.

\bibitem{Dalfovo_1999}
F.~Dalfovo, S.~Giorgini, L.~P. Pitaevskii, and S.~Stringari, ``Theory of
  bose-einstein condensation in trapped gases,'' {\em Rev. Mod. Phys.},
  vol.~71, pp.~463--512, Apr 1999.

\bibitem{greiner_2003}
M.~Greiner, C.~A. Regal, and D.~S. Jin, ``Emergence of a molecular
  bose--einstein condensate from a fermi gas,'' {\em Nature}, vol.~426,
  pp.~537--540, Dec 2003.

\bibitem{dyson_1967}
F.~J. Dyson and A.~Lenard, ``Stability of matter. i,'' {\em Journal of
  Mathematical Physics}, vol.~8, no.~3, pp.~423--434, 1967.

\bibitem{lenard_1968}
A.~Lenard and F.~J. Dyson, ``Stability of matter. ii,'' {\em Journal of
  Mathematical Physics}, vol.~9, no.~5, pp.~698--711, 1968.

\bibitem{pauskaukas_2001}
R.~Pa\ifmmode~\check{s}\else \v{s}\fi{}kauskas and L.~You, ``Quantum
  correlations in two-boson wave functions,'' {\em Phys. Rev. A}, vol.~64,
  p.~042310, Sep 2001.

\bibitem{plastino_2009}
A.~R. Plastino, D.~Manzano, and J.~S. Dehesa, ``Separability criteria and
  entanglement measures for pure states of n identical fermions,'' {\em {EPL}
  (Europhysics Letters)}, vol.~86, p.~20005, apr 2009.

\bibitem{Wootters_1997}
S.~A. Hill and W.~K. Wootters, ``Entanglement of a pair of quantum bits,'' {\em
  Phys. Rev. Lett.}, vol.~78, pp.~5022--5025, Jun 1997.

\bibitem{Rungta_2001}
P.~Rungta, V.~Bu\ifmmode~\check{z}\else \v{z}\fi{}ek, C.~M. Caves, M.~Hillery,
  and G.~J. Milburn, ``Universal state inversion and concurrence in arbitrary
  dimensions,'' {\em Phys. Rev. A}, vol.~64, p.~042315, Sep 2001.

\bibitem{manzano_2010}
D.~Manzano, A.~R. Plastino, J.~S. Dehesa, and T.~Koga, ``Quantum entanglement
  in two-electron atomic models,'' {\em Journal of Physics A: Mathematical and
  Theoretical}, vol.~43, p.~275301, jun 2010.

\bibitem{Zander_2012}
C.~{Zander}, A.~R. {Plastino}, M.~{Casas}, and A.~{Plastino}, ``{Entropic
  entanglement criteria for Fermion systems},'' {\em European Physical Journal
  D}, vol.~66, p.~14, Jan. 2012.

\bibitem{Uhlmann2010}
A.~{Uhlmann}, ``{Roofs and Convexity},'' {\em Entropy}, vol.~12,
  pp.~1799--1832, July 2010.

\bibitem{zurn_2012}
G.~Z\"urn, F.~Serwane, T.~Lompe, A.~N. Wenz, M.~G. Ries, J.~E. Bohn, and
  S.~Jochim, ``Fermionization of two distinguishable fermions,'' {\em Phys.
  Rev. Lett.}, vol.~108, p.~075303, Feb 2012.

\bibitem{girardeau_1960}
M.~Girardeau, ``Relationship between systems of impenetrable bosons and
  fermions in one dimension,'' {\em Journal of Mathematical Physics}, vol.~1,
  no.~6, pp.~516--523, 1960.

\bibitem{girardeau_2010}
M.~D. Girardeau, ``Two super-tonks-girardeau states of a trapped
  one-dimensional spinor fermi gas,'' {\em Phys. Rev. A}, vol.~82, p.~011607,
  Jul 2010.

\bibitem{rontani_2012}
M.~Rontani, ``Tunneling theory of two interacting atoms in a trap,'' {\em Phys.
  Rev. Lett.}, vol.~108, p.~115302, Mar 2012.

\bibitem{cuestas_2020_ferm}
E.~Cuestas, M.~D. Jiménez, and A.~P. Majtey, ``Entanglement and fermionization
  of two distinguishable fermions in a strict and non strict one-dimensional
  space,'' {\em Journal of Physics A: Mathematical and Theoretical}, 2020.

\bibitem{haldane_1991}
F.~D.~M. Haldane, ````fractional statistics'' in arbitrary dimensions: A
  generalization of the pauli principle,'' {\em Phys. Rev. Lett.}, vol.~67,
  pp.~937--940, Aug 1991.

\bibitem{wu_1994}
Y.-S. Wu, ``Statistical distribution for generalized ideal gas of
  fractional-statistics particles,'' {\em Phys. Rev. Lett.}, vol.~73,
  pp.~922--925, Aug 1994.

\bibitem{guan_2007}
X.~W. Guan, M.~T. Batchelor, C.~Lee, and M.~Bortz, ``Phase transitions and
  pairing signature in strongly attractive fermi atomic gases,'' {\em Phys.
  Rev. B}, vol.~76, p.~085120, Aug 2007.

\bibitem{shu_2010}
S.~Chen, X.-W. Guan, X.~Yin, L.~Guan, and M.~T. Batchelor, ``Realization of
  effective super tonks-girardeau gases via strongly attractive one-dimensional
  fermi gases,'' {\em Phys. Rev. A}, vol.~81, p.~031608, Mar 2010.

\bibitem{levin_2006}
M.~Levin and X.-G. Wen, ``Detecting topological order in a ground state wave
  function,'' {\em Phys. Rev. Lett.}, vol.~96, p.~110405, Mar 2006.

\bibitem{kitaev_2006}
A.~Kitaev and J.~Preskill, ``Topological entanglement entropy,'' {\em Phys.
  Rev. Lett.}, vol.~96, p.~110404, Mar 2006.

\bibitem{bonderson_2017}
P.~Bonderson, C.~Knapp, and K.~Patel, ``Anyonic entanglement and topological
  entanglement entropy,'' {\em Annals of Physics}, vol.~385, pp.~399--468,
  2017.

\bibitem{santachiara_2007}
R.~Santachiara, F.~Stauffer, and D.~C. Cabra, ``Entanglement properties and
  momentum distributions of hard-core anyons on a ring,'' {\em Journal of
  Statistical Mechanics: Theory and Experiment}, vol.~2007, p.~L05003, may
  2007.

\bibitem{mani_2020}
H.~S. Mani, R.~N, and V.~V. Sreedhar, ``Quantum entanglement in one-dimensional
  anyons,'' {\em Phys. Rev. A}, vol.~101, p.~022314, Feb 2020.

\bibitem{lofranco_2016}
R.~Lo~Franco and G.~Compagno, ``Quantum entanglement of identical particles by
  standard information-theoretic notions,'' {\em Scientific Reports}, vol.~6,
  p.~20603, Feb 2016.

\bibitem{compagno_2018}
G.~Compagno, A.~Castellini, and R.~Lo~Franco, ``Dealing with indistinguishable
  particles and their entanglement,'' {\em Philosophical Transactions of the
  Royal Society A: Mathematical, Physical and Engineering Sciences}, vol.~376,
  no.~2123, p.~20170317, 2018.

\bibitem{lourenco_2019}
A.~C. Louren\ifmmode~\mbox{\c{c}}\else \c{c}\fi{}o, T.~Debarba, and E.~I.
  Duzzioni, ``Entanglement of indistinguishable particles: A comparative
  study,'' {\em Phys. Rev. A}, vol.~99, p.~012341, Jan 2019.

\end{thebibliography}

\end{document}